\documentclass[twocolumn]{aastex63}

\usepackage{upgreek}
\usepackage{textcomp}
\usepackage{wasysym}
\usepackage{blindtext}
\usepackage{lineno}

\shorttitle{Extreme T Subdwarf Candidate}
\shortauthors{Hunter Brooks}

\graphicspath{{./}{figures/}}

\begin{document}

\title{Discovery of CWISE J052306.42-015355.4, an Extreme T Subdwarf Candidate}

\author[0000-0002-5253-0383]{Hunter Brooks}
\affiliation{Department of Astronomy and Planetary Science, Northern Arizona University, Flagstaff, AZ 86011, USA}

\author[0000-0003-4269-260X]{J.\ Davy Kirkpatrick}
\affiliation{IPAC, Mail Code 100-22, Caltech, 1200 E. California Blvd., Pasadena, CA 91125, USA}

\author[0000-0001-7896-5791]{Dan Caselden}
\affiliation{Backyard Worlds: Planet 9, USA}

\author[0000-0002-6294-5937]{Adam C. Schneider}
\affiliation{United States Naval Observatory, Flagstaff Station, 10391 West Naval Observatory Rd., Flagstaff, AZ 86005, USA} 
\affiliation{Department of Physics and Astronomy, George Mason University, MS3F3, 4400 University Drive, Fairfax, VA 22030, USA }

\author[0000-0002-1125-7384]{Aaron M. Meisner}
\affiliation{NSF's National Optical-Infrared Astronomy Research
Laboratory, 950 N. Cherry Ave., Tucson, AZ 85719, USA}

\author[0000-0001-6251-0573]{Jacqueline K. Faherty}
\affiliation{Department of Astrophysics, American Museum of Natural History, Central Park West at 79th Street, New York, NY 10024, USA}

\author[0000-0003-2478-0120]{ S.L.Casewell}
\affiliation{School of Physics and Astronomy, University of Leicester, University Road, Leicester, LE1 7RH, UK}

\author[0000-0002-2387-5489]{Marc J. Kuchner}
\affiliation{NASA Goddard Space Flight Center, Exoplanets and Stellar Astrophysics Laboratory, Code 667, Greenbelt, MD 20771, USA}

\author{The Backyard Worlds:  Planet 9 Collaboration}

\begin{abstract}
We present the discovery of CWISE J052306.42$-$015355.4, which was found as a faint, significant proper motion object (0.52 $\pm$ 0.08 arcsec yr$^{-1}$) using machine learning tools on the unWISE re-processing on time series images from the {\it Wide-field Infrared Survey Explorer}. Using the CatWISE2020 W1 and W2 magnitudes along with a $J-$band detection from the VISTA Hemisphere Survey, the location of CWISE J052306.42$-$015355.4 on the W1$-$W2 vs. $J-$W2 diagram best matches that of other known, or suspected, extreme T subdwarfs. As there is currently very little knowledge concerning extreme T subdwarfs we estimate a rough distance of $\le$ 68 pc, which results in a tangential velocity of $\le$ 167 km s$^{-1}$, both of which are tentative. A measured parallax is greatly needed to test these values. We also estimate a metallicity of $-1.5 <$ [M/H] $< -0.5$ using theoretical predictions.
\end{abstract}


\section{Introduction}
Data from NASA's {\it Wide-field Infrared Survey Explorer} (hereafter, {\it WISE}; \citealt{Wright et al. 2010}) and {\it Near-Earth Object WISE} (hereafter, {\it NEOWISE}; \citealt{Mainzer et al. 2011}) missions have made searches for nearby, low-temperature objects considerably easier. The long time baseline between the earliest {\it WISE} data in early 2010 and the most recent (ongoing) {\it NEOWISE} data means that objects with small proper motion can now be detected using only {\it WISE} data. The first {\it WISE} searches for nearby objects were reliant mainly on colors (\citealt{Kirkpatrick et al. 2011}), but the addition of motion information has enabled the discovery of brown dwarfs with unusual colors, a subset of which are believed to be old and metal-poor (\citealt{Schneider et al. 2020, Meisner et al. 2020, Meisner et al. 2021}, and \citealt{Kirkpatrick et al. 2021b}).

Several tools have been developed to exploit these motions. First, the CatWISE2020 Catalog (\citealt{Marocco et al. 2021}) uses a long, 8-yr time baseline between {\it WISE} and {\it NEOWISE} data to compute proper motion measurements for all detected objects. Second, the WiseView user interface (\citealt{Caselden et al. 2018}), which was designed to aid the Backyard Worlds: Planet 9 citizen science group (hereafter, BYW; \citealt{Kuchner et al. 2017}), allows for visual confirmation of motion objects. It is particularly useful for those citizens doing their own targeted color or motion searches on {\it WISE} data products. Third, machine learning algorithms applied to the stack of epochal unWISE coadds (\citealt{Lang(2014)}) from  \cite{Meisner et al. (2018)} can also be trained to identify detections exhibiting substantial proper motion.

All three methods have been used in the discovery and subsequent classification of the object CWISE J052306.42$-$015355.4 (hereafter, J0523$-$0153), a new esdT candidate. In \S\ref{Section two} we discuss the discovery of the object. Below we further describe or estimate other attributes of J0523$-$0153, including photometry (\S\ref{photometry}), spectral type (\S\ref{spectra}), distance (\S\ref{distance}), velocity (\S\ref{movement}), and metallicity (\S\ref{metallicity}). \S\ref{dis} addresses different esdT candidates and compares them to J0523$-$0153. Our conclusions are summarized in \S\ref{con}.

\break

\section{History of T Subdwarfs and Extreme T Subdwarfs}
T subdwarfs (hereafter,  sdT) are a low-metallicity counterparts of regular T dwarfs (hereafter, dT), while extreme T subdwarfs (hereafter, esdT) are those with even lower metallicity. The first suspected sdT candidate is 2MASS J09373487+2931409 (\citealt{Burgasser et al. 2002}). However, the first confirmed sdT is WISE J200520.38+542433.9 (\citealt{Mace et al. 2013}).  Its colors and near-infrared spectrum were unusual, and its common proper motion with a low metallicity primary, Wolf 1130AB, provided the explanation. This paved the way to the discovery of other sdTs. 

Thereafter, two more sdTs were discovered by \cite{Pinfield et al. 2014}, WISE J001354.39+063448.1 and WISE J083337.82+005214.1, both of which were identified through their unusual colors and high proper motions. \cite{Burningham et al. 2014} discovered the sdT6.5 ULAS J131610.28+075553.0, which was found to be a potential halo object. These known sdTs all have suppressed $K-$band flux compared to dTs. \cite{Zhang et al. 2019} selected 41 sdT5-9 from the literature by this feature in their $K-$band spectra, and discussed how sdTs differ from dTs and L subdwarfs in their characteristics. 

Two other spectroscopically unusual T dwarfs, WISE J041451.67$-$585456.7 (hereafter, J0414$-$5854) and WISE J181006.18$-$101000.5 (hereafter, J1810$-$1010), were then discovered by \cite{Schneider et al. 2020}. These two objects have distinctive $J-K$ and $J-$W2 colors from known dTs, sdTs, or sdLs. They are in the same metallicity range as L extreme subdwarfs (in Table 8 of \citealt{Zhang et al. 2017}), but cooler than esdLs. This suggesting that they are esdTs. Later, in \cite{Meisner et al. 2021} two new esdT candidates were discovered, WISEA J041451.67$-$585456.7 and WISEA J181006.18$-$101000.5, which had similar colors to the first esdTs which identified them as candidates.

\section{Discovery of J0523\textendash0153}\label{Section two}

\begin{figure}[ht]
        \centering
        \includegraphics[scale = 0.21]{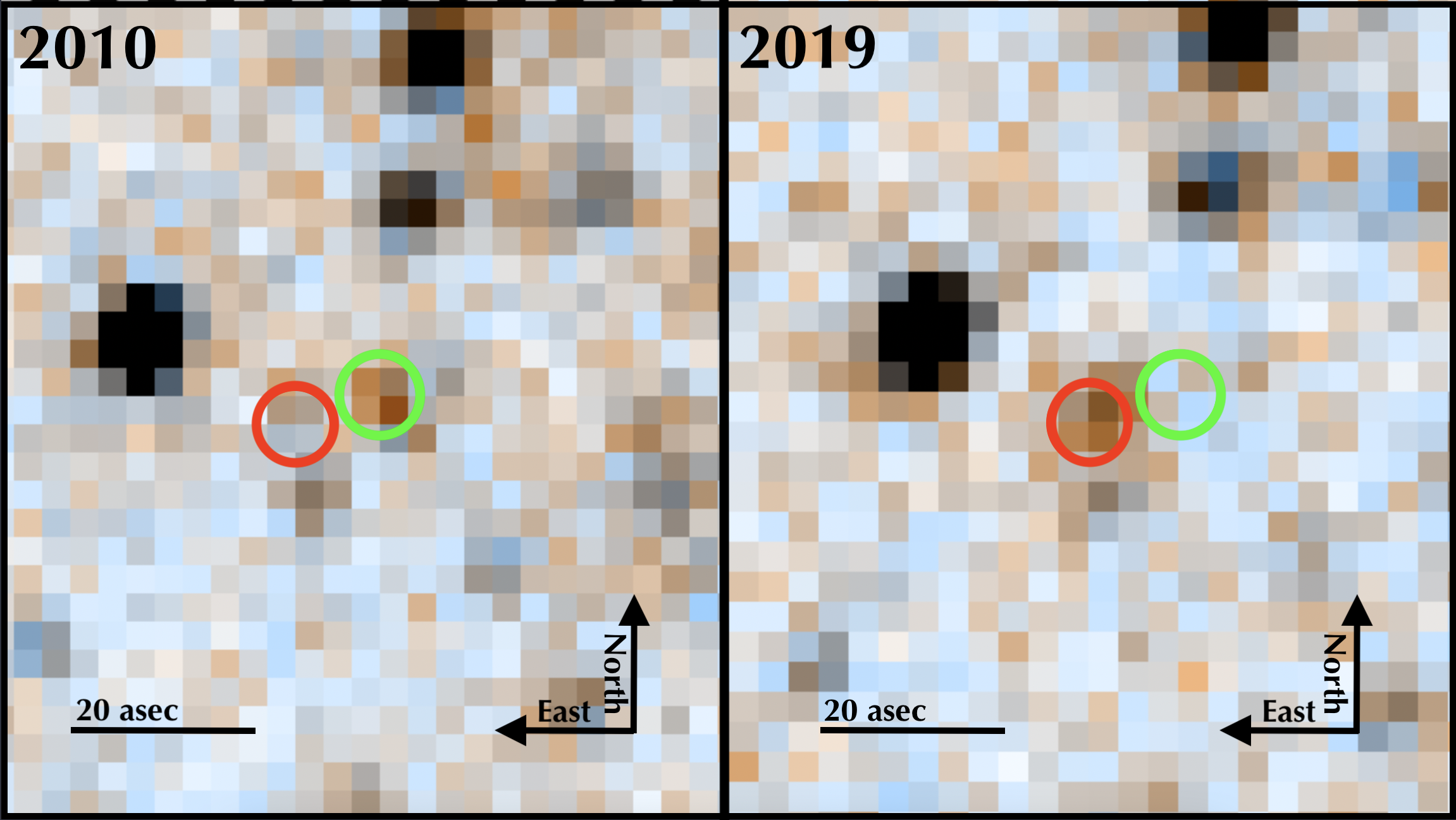}
        \caption{The movement of J0523$-$0153 from $WISE$ in 2010 through $NEOWISE$ in 2019. The green circle shows the position of J0523$-$0153 at 2010, and the red circle shows it at 2019. }
        \label{Figure 1}
\end{figure}

J0523$-$0153 was recovered by our team in multiple ways over the course of our extensive searches for new members of the solar neighborhood. Co-author Dan Caselden designed a machine learning algorithm, nicknamed SMDET (\citealt{Caselden et al. 2020}), to aid in the discovery of motion objects across unWISE epochal coadds (\citealt{Meisner et al. (2018)}). SMDET searches for flux groupings whose pixel positions move across the time sequence of coadds. SMDET is trained on synthetic motion objects added to the unWISE coadds. In addition, SMDET also produces a computer-generated mask showing the general movement of the object across the image, to aid the eye in finding the object in question. Author Hunter Brooks discovered J0523$-$0153 using SMDET's list of candidate motion objects. Figure \ref{Figure 1} shows the movement of J0523$-$0153 from $WISE$ in 2010 through $NEOWISE$ in 2019. This object was also submitted as a motion discovery to BYW by citizen scientist Sam Goodman, using tools provided and developed for that Zooniverse project. Additionally, J0523$-$0153 was found to have significant motion in the CatWISE2020 Catalog. We list the proper motion measurements along with the W1 and W2 magnitudes in Table 1.

\section{Characteristics of J0523\textendash0153}

\subsection{Photometry}\label{photometry}

\begin{figure}[ht]
        \centering
        \includegraphics[scale = 0.37]{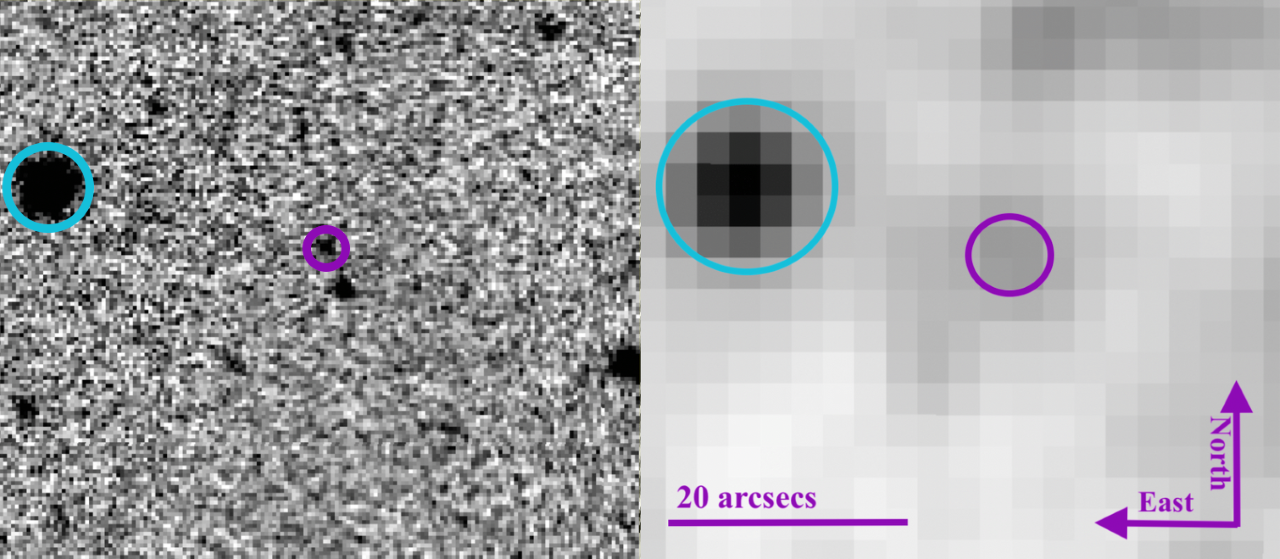}
        \caption{VISTA Hemisphere Survey (DR6) $J-$band image is on the left alongside $WISE$ W2-band image on the right. Both figures show the same patch of sky at the same scale, which is indicated on the right-hand panel. The purple circle is J0523$-$0153 and the blue circle is a near-by object that can be used as a reference. }
        \label{Figure 2}
\end{figure}

The CatWISE2020 W1 and W2 magnitudes for J0523$-$0153 are 17.27$\pm$0.06 mag and 15.91$\pm$0.06 mag, respectively, resulting in a color of W1$-$W2 = 1.36$\pm$0.08 mag. This object was not detected in the 2MASS survey (\citealt{Skrutskie et al. 2006}). We then searched the VISTA Science Archive (\citealt{Cross et al. (2012)}) and WFCAM Science Archive (\citealt{Hambly et al. (2008)}). A deep $J-$band detection was found on the VISTA Hemisphere Survey as shown in Figure \ref{Figure 2}, with $J_{MKO}$ = 19.75 $\pm$ 0.186 (mag).

\subsection{Spectral Type}\label{spectra}
\begin{figure*}[ht]
    \centering
    \includegraphics[scale = 0.5]{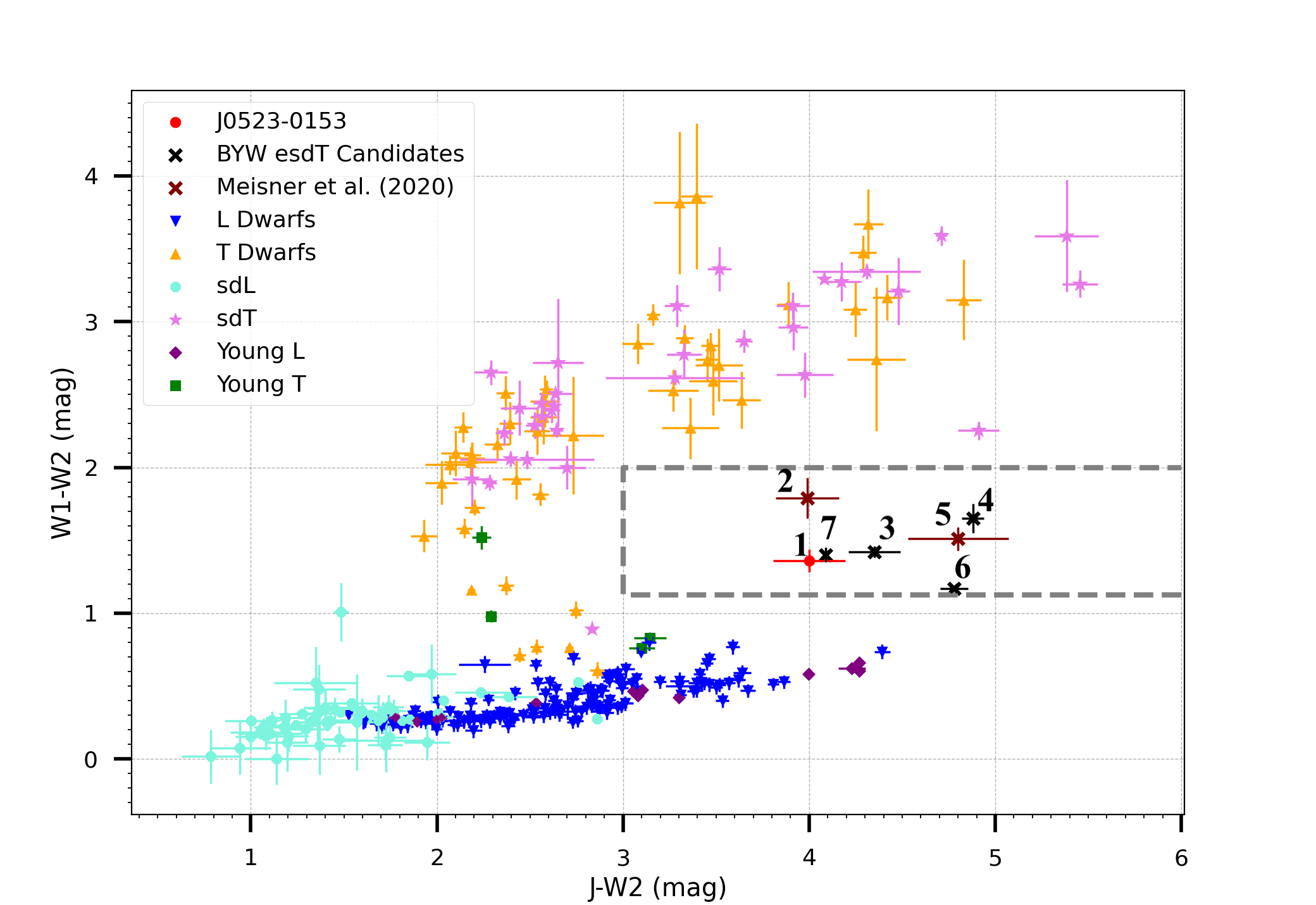}
    \caption{Color-color diagram adapted from Figure 1 of \cite{Meisner et al. 2021} and \cite{Kirkpatrick et al. 2016}. The gray box was discussed in \cite{Meisner et al. 2021} to contain all known esdTs, and candidates; therefore, we call it the esdT box. The location of J0523$-$0153 (red circle) places it well within the esdT box. L dwarfs (blue downward triangles), T dwarfs (orange upward triangles), young L (purple diamonds), young T dwarfs (green squares), L subdwarfs (light blue circles), and T subdwarfs (pink stars) from the 20-pc L, T, and Y dwarf sample of \cite{Kirkpatrick et al. 2021a} are shown for comparison, as are the extreme T subdwarf candidates, from BYW papers (black crosses), of \cite{Schneider et al. 2020} and \cite{Meisner et al. 2021}. WISEA J153429.75$-$104303.3 (\citealt{Kirkpatrick et al. 2021b}) has a large $J-$W2 color, falling at a $J-$W2 value that is too red for this graph. Both CWISEP J050521.29\textendash591311.7 and CWISEP J090536.35+740009.1 from \cite{Meisner et al. 2020} are shown by maroon crosses. Objects in the esdT box are: 1: J0523$-$0153, 2: J0905+7400, 3: J0414$-$5854, 4: J2217$-$1454, 5: J0505$-$5913, 6: J1810$-$1010, 7: J0738$-$6643 (all of these objects can be found in Table 2). }
    \label{Figure 3}
\end{figure*}

With the W1$-$W2 color and $J-$W2 color, we can plot J0523$-$0153 on a color-color plot like that of Figure 8 of \cite{Kirkpatrick et al. 2016}. This is illustrated in Figure~\ref{Figure 3}, showing that the $J-$W2 color is extremely red for its W1$-$W2 color, compared to the location of normal brown dwarfs. Such colors imply an extreme T subdwarf as defined by \cite{Meisner et al. 2021}. As a result of there being no spectral type vs. color relations for esdTs yet, we approximate a mid-T type using the relation for normal T dwarfs, as done in \cite{Meisner et al. 2021} and \cite{Schneider et al. 2020}. We provide more context with other known and suspected T subdwarfs in \S\ref{Close objects}.

On the other hand, when we look at Figure 8(i) from \cite{Zhang et al. 2019} at a fixed spectral type, sdTs are redder in W1$-$W2 compared to normal T dwarfs. If we assume a similar behavior for esdTs then we would expect J0523$-$0153 to be an early esdT. This is also supported by WISEA J041451.67$-$585456.7 (hereafter, J0414$-$5854; \citealt{Schneider et al. 2020}), which has similar colors (Table 2 and Figure~\ref{Figure 3}) to J0523$-$0513, has a tentative spectroscopic classification of esdT0$\pm$1. Thus, a slightly preferred estimation for the spectral type of J0523$-$0512 is $\sim$esdT0.

\begin{table}[ht]

    \centering
    
    \caption{Compiled Data for J0523$-$0153}
    \begin{tabular}{l c r} 

        \hline
        \hline
        Parameter & Value & Ref. \\
        \hline
        \hline
         & Astrometry & \\
         \hline
        RA & 80.7767549 deg & 2\\
        DEC & $-$1.8987341 deg & 2\\
        $\mu_{RA}$ & $0.52{\pm}0.05$ arcsec yr$^{-1}$ & 2 \\
        $\mu_{Dec}$ & $-0.04{\pm}0.06$ arcsec yr$^{-1}$ & 2 \\
        $\mu_{total}$ & $0.52{\pm}0.08$ arcsec yr$^{-1}$ & 1 \\
        \hline
         & Photometry & \\
         \hline
        $J_{MKO}$ & 19.75$\pm$0.186 mag & 1,3 \\
        W1 & 17.27$\pm$0.06 mag & 2\\
        W2 & 15.91$\pm$0.06 mag & 2\\
        $J-$W2 & 3.84$\pm$0.195 mag & 1\\
        W1$-$W2 & 1.36$\pm$0.08 mag & 2\\
        \hline
         & Derived Quantities & \\
         \hline
        distance & $\le$ 68 pc & 1 \\
        $V_{tan}$ & $\le$ 167 km s$^{-1}$  & 1 \\
        \hline
        
    \end{tabular}
    \footnotesize{Reference Codes: (1) This publication, (2) CatWISE2020 Catalog; \cite{Marocco et al. 2021}, (3) VISTA Science Archive; \cite{Cross et al. (2012)}}
\end{table}
\normalsize

\break

\subsection{Distance}\label{distance}
As there is no confirmed esdT that has a measured trigonometric parallax, we have no clear way to estimate the distance. In Figure 7 of \cite{Meisner et al. 2021} the LOWZ models show that W1$-$W2 is not as temperature sensitive as $J-$W2 at colors similar to esdTs. A better indicator of temperature, or spectral type, would be obtained from the $J-$W2 color. Comparing our value of $J-$W2 3.84$\pm$0.195 mag to Figure 20(c) of \cite{Kirkpatrick et al. 2021a}, we obtain M$_{W2}$ $\approx$ 14.0 using Figure 15 of the same paper, where they equate $Spitzer$ ch2 photometry with $WISE$ W2 photometry, which results in a distance of $\approx$ 24 pc.

If we use the standard W1$-$W2 vs. spectral type relation of \cite{Kirkpatrick et al. 2021a}, which applies only to solar metallicity brown dwarfs, we estimate a T5 spectral type. From Figure 16(d) of \cite{Kirkpatrick et al. 2021a} we find that a regular T5 has an M$_{W2}$ $\approx$ 12.5  which results in a distance of $\approx$ 48 pc. However, it is much more likely that J0523$-$0153 is an esdT0 therefore, using the same process as before we get an M$_{W2}$ $\approx$ 11.75  which results in a distance of $\sim$68 pc.

Another way to estimate the distance of J0523$-$0153 is looking at Figure 20(c) of \cite{Kirkpatrick et al. 2021a}. We see that sdTs tend to be brighter in $M_{ch2}$. If we plot J0523$-$0153 brighter, at about the distance that sdTs are removed from the regular sequence, we estimate $M_{ch2}$ $\approx$ 13. This results in a distance of $\approx$ 38 pc.

We can also estimate other distance limits using the measured proper motion. As \cite{Kirkpatrick et al. 2021b} shows, the highest velocity of any brown dwarf in the 20 pc sample is $\sim$200 km s$^{-1}$. If we instead use this value as a reasonable upper bound, then we find that J0523$-$0153 most likely has a distance \textless \space 84 pc. Even though this is a very loose constraint, it is in agreement with our previous estimates.

As a result of the total velocity distance estimate being a rough upper bound, we revert to our previous estimate for esdT0's of $\le$ 68 pc for J0523$-$0153. A measured trigonometric parallax is needed to better understand this, and other, esdT candidates.

\subsection{Proper Motion and Tangential Velocity}\label{movement}
J0523$-$0153 is shown in Figure~\ref{Figure 1} to have  significant proper motion, and this is measured in the CatWISE2020 Catalog as $\mu_{RA} = 0.52{\pm}0.05$ arcsec yr$^{-1}$, $\mu_{Dec} = -0.04{\pm}0.06$ arcsec yr$^{-1}$, and $\mu_{tot} = 0.52{\pm}0.08$ arcsec yr$^{-1}$. Using our poorly constrained distance estimate of $\le$ 68 pc, we find a tangential velocity of $\le$ 167 km s$^{-1}$.

\begin{table*}[ht]

    \centering
    
    \caption{Compiled Data for all confirmed or candidate esdTs}
    \begin{tabular}{l c c c c c c r} 

        \hline
        \hline
         & & & Photometry & & & \\
        \hline
        \hline
        Object Name & W1 (mag) & W2 (mag) & J (mag) & W1$-$W1 (mag) & $J-$W2 (mag) & Ref. \\
        \hline
                
        J0414$-$5854 & 16.71 $\pm$ 0.03 & 15.29 $\pm$ 0.02 & 19.63 $\pm$ 0.11 & 1.42 $\pm$ 0.04 & 4.35 $\pm$ 0.14 & 2 \\
        
        J0505$-$5913 & 17.64 $\pm$ 0.06 & 16.13 $\pm$ 0.05 & 20.93 $\pm$ 0.29 & 1.51 $\pm$ 0.08 & 4.80 $\pm$ 0.27 & 3 \\

        J0523$-$0153 & 17.27 $\pm$ 0.06 & 15.91 $\pm$ 0.06 & 19.75$\pm$0.186 & 1.36 $\pm$ 0.08 & 3.84$\pm$0.195 & 1 \\
        
        J0738$-$6643 & 17.22 $\pm$ 0.04 & 15.82 $\pm$ 0.04 & \textgreater \space 19.92 & 1.40 $\pm$ 0.05 & \textgreater \space 4.09 & 4 \\
        
        J0905+7400 & 18.25 $\pm$ 0.16 & 16.46 $\pm$ 0.09 & 20.45 $\pm$ 0.14 & 1.79 $\pm$ 0.14 & 3.99 $\pm$ 0.17 & 3 \\
        
        J1534$-$1043 & 18.18 $\pm$ 0.19 & 16.15 $\pm$ 0.08 & \textgreater \space 23.80 & 2.04 $\pm$ 0.21 & \textgreater \space 7.66 & 5 \\
        
        J1810$-$1010 & 13.65 $\pm$ 0.02 & 12.48 $\pm$ 0.01 & 17.26 $\pm$ 0.02 & 1.17 $\pm$ 0.02 & 4.78 $\pm$ 0.02 & 2 \\
        
        J2217$-$1454 & 17.43 $\pm$ 0.08  & 15.78 $\pm$ 0.06 & 20.66 $\pm$ 0.02 & 1.65 $\pm$ 0.10 & 4.88 $\pm$ 0.06 & 4 \\
    
        \hline
        \hline
         & & & Astrometry & & & & \\
        \hline
        \hline
         Object Name & & $\mu_{RA}$ (arcsec yr$^{-1}$) & $\mu_{Dec}$ (arcsec yr$^{-1}$) & $\mu_{total}$ (arcsec yr$^{-1}$) & & Ref. \\
        \hline
        
        J0414$-$5854 & & 0.21 $\pm$ 0.03 & 0.65 $\pm$ 0.03 &  0.68 $\pm$ 0.03 & & 2 \\
        
        J0505$-$5913 & & 0.46 $\pm$ 0.04 & -1.00 $\pm$ 0.04 &  1.10 $\pm$ 0.04 & & 3 \\

        J0523$-$0153 & & 0.52 $\pm$ 0.05 & $-$0.04 $\pm$ 0.06 & 0.52 $\pm$ 0.08 & & 1 \\
        
        J0738$-$6643 & & 0.77 $\pm$ 0.03 & $-$0.43 $\pm$ 0.03 & 0.88 $\pm$ 0.03 & & 4 \\
        
        J0905+7400 & & 0.47 $\pm$ 0.07 & -1.49 $\pm$ 0.07 & 1.56 $\pm$ 0.07 & & 3 \\
        
        J1534$-$1043 & & $-$1.25 $\pm$ 0.01 & $-$2.38 $\pm$ 0.01 & 2.69 $\pm$ 0.01 & & 5 \\
        
        J1810$-$1010 & & $-$1.13 $\pm$ 0.01 & $-$0.21 $\pm$ 0.01 & 1.16 $\pm$ 0.01  & & 2 \\
        
        J2217$-$1454 & & 1.64 $\pm$ 0.07 & $-$0.92 $\pm$ 0.06 & 1.88 $\pm$ 0.07 & & 4 \\
    
        \hline
        \hline
         & & & Derived Quantities & & & \\
        \hline
        \hline
        Object Name & Distance (pc) & $V_{tan}$ (km s$^{-1}$) & Metallicity (dex) & $T_{\rm eff}$ (K) & Mass ($M_{\astrosun}$) & Ref.\\
        \hline
        
        J0414$-$5854 & 52$-$94 & 170$-$307 & $\approx$ $-$1 [Fe/H] & 1300 $\pm$ 100 & 0.075$-$0.080 & 2 \\
        
        J0505$-$5913 & 34.9$-$46.3 & 181$-$243 & ... & 846 $\pm$ 88 & ... & 3 \\

        J0523$-$0153 & $\le$ 68 & $\le$ 167 & $-$1.5 to $-$0.5 [M/H] & ... & ... & 1 \\
        
        J0738$-$6643 & 35.2$-$54 & 145$-$225 & ... & ... & ... & 4 \\
        
        J0905+7400 & 41.6$-$55.1 & 305$-$410 & ... & 930 $\pm$ 87 & ... & 3 \\
        
        J1534$-$1043 & 15.1$-$17.7 & 191.5$-$223.4 & $\approx$ $-$1.5 [M/H] & $-$500 & ... & 5 \\
        
        J1810$-$1010 & 14$-$67 & 77$-$360 & $\le$ $-$1 [Fe/H] & 1300 $\pm$ 100 & 0.075$-$0.080 & 2 \\

        J2217$-$1454 & 31.6$-$48.9 & 282$-$436 & ... & ... & ... & 4 \\

        \hline 
    \end{tabular}
    \footnotesize{Reference Codes: (1) This publication, (2) \cite{Schneider et al. 2020}, (3) \cite{Meisner et al. 2020}, (4) \cite{Meisner et al. 2021}, (5) \cite{Kirkpatrick et al. 2021b}}. Note that only one esdT has a measured parallax, J1534$-$1043. For the others, the distances are estimates. This results in the $V_{tan}$ also being a rough estimate for every object other than J1534$-$1043.
\end{table*}

\break

\subsection{Metallicity}\label{metallicity}
Based on the location of J0523$-$0153 in the color-color plot of Figure~\ref{Figure 3}, we can attempt to deduce its metallicity. If we plot it on Figure 7 of \cite{Meisner et al. 2021}, which is similar to Figure~\ref{Figure 3} here, we find that it falls along the LOWZ model with metallicities of [M/H] = $-$2. However, WISEA J153429.75$-$104303.3 (hereafter, J1534$-$1043; \citealt{Meisner et al. 2020} and \citealt{Kirkpatrick et al. 2021b}) falls in a far more unusual spot on the color-color diagram and is believed to have a metallicity [M/H] $\approx$ $-$1.5 (\citealt{Kirkpatrick et al. 2021b}), more metal rich than the LOWZ model predicts.

For additional guidance, we can examine the location of WISE J200520.38+542433.9 (hereafter, J2005+5424; \citealt{Mace et al. 2013}) on Figure~\ref{Figure 3}, since this object is a co-moving companion to a star with a known subsolar metallicity of [Fe/H] = $-$0.64 $\pm$ 0.17 dex. If we use the LOWZ model, it predicts a subsolar metallicity of [M/H] = $-$1.0. However, the actual measured metallicity from the primary of the system is higher than the prediction adopting the LOWZ model. Thus, the models are making predictions that are systematically too metal poor. The location of J2005+5424 on Figure~\ref{Figure 3} is closer to the normal T dwarf sequence than that of J0523$-$0153, so this means that this object would have a higher metallicity. But J1534$-$1043 is further removed from the normal sequence than both of these objects. Since we believe that J0523$-$0153 is less metal poor than J1534$-$1043 and more than J2005+5424, then it most likely has a metallicity in the range of $-$1.5 \textless \space [M/H] \textless \space $-$0.5.

\section{Discussion}\label{dis}  
There are eight other candidates or confirmed esdTs so far published (Table 2), so it is useful to compare J0523$-$0153 to these esdTs. Most esdTs have been discovered and discussed in papers by the BYW team (\citealt{Schneider et al. 2020}, \citealt{Meisner et al. 2021}, and \citealt{Kirkpatrick et al. 2021b}). There are two esdT candidates found and discussed in \cite{Meisner et al. 2020} that are not from the BYW community. Since this family of brown dwarfs is relatively new and exceedingly faint, the follow-up data, which are hard to come by, are sparse.

\subsection{Known esdTs and other esdT Candidates}\label{Close objects}
The first extreme T subdwarf candidates were discussed in \cite{Schneider et al. 2020}. Both J0414$-$5854 and J1810$-$1010 have spectroscopic signatures indicative of sub-solar metallicity, which identified them as T subdwarfs. The authors came to this conclusion using their Figure 2, which compares the spectra of these objects to the spectra of known dTs and sdTs. Accordingly, their Figure 4, which compares the spectra to model predictions in both temperature and metallicity.

J0414$-$5854 and J1810$-$1010 were both shown to have similar unusual colors compared to known brown dwarfs, as shown in Table 2. As shown in Figure 3 of \cite{Schneider et al. 2020} middle to late esdTs are bluer in W1$-$W2 while redder in $J-$W2, $J-$K, and $J-$H when compared to regular dTs.

Both CWISEP J050521.29$-$591311.7 and CWISEP J090536.35+740009.1 (hereafter, J0505$-$5913 and J0905+7400) were discussed in \cite{Meisner et al. 2020} as mid-T dwarf candidates, even possibly T subdwarfs. We would now list both objects as mid-esdT candidates based on their colors. These magnitudes, found in Table 2, are very similar to other esdTs that are found in the esdT box in \cite{Meisner et al. 2021}, shown in Figure~\ref{Figure 3}. They estimated that J0505$-$5913 is a T6.5 and J0905+7400 is a T6, assuming both are normal T-type dwarfs. With ch1$-$ch2 data from $Spitzer$, they find distance estimates for J0505$-$5913 and J0905+7400 of 34.9$-$46.3 pc and 41.6$-$55.1 pc, respectively.

\cite{Meisner et al. 2021} discussed two new esdT candidates, CWISE J073844.52$-$664334.6 (hereafter, J0738$-$6643) and CWISE J221706.28$-$145437.6 (hereafter, J2217$-$1454). J0738$-$6643 was found to have characteristics of a T5$\pm$1.2. J2217$-$1454 has a spectral type estimate of T5.5$\pm$1.2, these estimates came from the W1$-$W2 vs. brown dwarf spectral type, which has been shown to be problematic. This is similar to J0738$-$6643 and comparable to J0523$-$0153, a reminder that these values are poor estimates given that these objects are likely not normal-T type dwarfs. 

The most exotic of the extreme subdwarf candidates is WISEA J153429.75$-$104303.3 (hereafter, J1534$-$1043; \citealt{Meisner et al. 2020} and \citealt{Kirkpatrick et al. 2021b}). J1534$-$1043 was so faint in near-infrared that those authors needed to use the $Hubble$ $Space$ $Telescope$ to obtain a detection of it. Even with this hurdle, those authors show that J1534$-$1043 is the reddest of all esdTs in $J-$W2, including candidates, and is possibly an esdY. Even though it has a red $J-$W2, it still falls within the W1$-$W2 color constraint of the esdT box of Figure~\ref{Figure 3}. In \S\ref{metallicity} we discussed how this object compares to J0523$-$0153 in metallicity. There is a temperature estimate of $T_{\rm eff}$ \space \textless \space 500 K using the theoretical LOWZ model, but these predictions have not been tested yet.

\subsection{Implications on J0523\textendash0153}
When looking at J0523$-$0153 we can see that it has many neighbors in Figure~\ref{Figure 3}. Peering at confirmed/candidate esdTs that have similar colors to J0523$-$0153, we can conclude that they likely have similar characteristics. For example, as both J0738$-$6653 and J0523$-$0153 have similar colors, we can assume they have comparable spectral types. Likewise, the metallicities that we estimated are comparable to similarly colored esdTs that have spectra, with which a more accurate metallicity can be estimated. 

\section{Conclusion} \label{con}
We have announced the discovery of the proper motion object J0523$-$0153 found in {\it WISE} data. We estimate that the object is an early esdT dwarf candidate. With our current models, we provide a crude estimate for both a distance and tangential velocity. Applying different methods for distance estimates, we find $\le$ 68 pc. Additionally, using proper motion values from the CatWISE2020 Catalog and the distance estimate stated before, we calculate a tangential velocity of $\le$ 167 km s$^{-1}$. Using its placement on color-color diagrams, we estimate the metallicity to fall in the range $-$1.5 to $-$0.5 dex. Follow-up photometry, spectroscopy, and a trigonometric parallax are needed to help further reveal the properties of J0523$-$0153.
    
\section{Acknowledgements} 
Backyard Worlds research was supported by NASA grant 2017-ADAP17-0067 and by the NSF under grants AST- 2007068, AST-2009177, and AST-2009136. SLC is supported by an STFC Ernest Rutherford Research Fellowship. We want to thank the Student Astrophysics Society\footnote{\url{https://www.studentastrophysicssociety.com}} for providing the resources that enabled the pairing of high school and undergraduate students with practicing astronomers and advanced citizen scientists. This work makes use of data products from WISE/NEOWISE, which is a joint project of UCLA and JPL/Caltech, funding by NASA. The CatWISE effort is led by the Jet Propulsion Laboratory, California Institute of Technology, with funding from NASA’s Astrophysics Data Analysis Program. We would like to thank our anonymous referee for the suggestions that improved the paper. 

\software{WiseView (\citealt{Caselden et al. 2018}).}


\begin{thebibliography}{}

\bibitem[Burgasser et al.(2002)]{Burgasser et al. 2002} Burgasser, A.~J., Kirkpatrick, J.~D., Brown, M.~E., et al.\ 2002, \apj, 564, 421. doi:10.1086/324033

\bibitem[Burgasser et al.(2005)]{Burgasser et al. 2005} Burgasser, A.~J., Kirkpatrick, J.~D., \& L{\'e}pine, S.\ 2005, 13th Cambridge Workshop on Cool Stars, Stellar Systems and the Sun, 560, 237

\bibitem[Burgasser(2008)]{Burgasser et al. 2008} Burgasser, A.~J.\ 2008, 14th Cambridge Workshop on Cool Stars, Stellar Systems, and the Sun, 384, 126

\bibitem[Burningham et al.(2014)]{Burningham et al. 2014} Burningham, B., Smith, L., Cardoso, C.~V., et al.\ 2014, \mnras, 440, 359. doi:10.1093/mnras/stu184

\bibitem[Caselden et al.(2018)]{Caselden et al. 2018} Caselden, D., Westin, P., Meisner, A., et al.\ 2018, Astrophysics Source Code Library. ascl:1806.004

\bibitem[Caselden et al.(2020)]{Caselden et al. 2020} Caselden, D., Colin, G., Lack, L., et al.\ 2020, \aas

\bibitem[Cross et al.(2012)]{Cross et al. (2012)} Cross, N.~J.~G., Collins, R.~S., Mann, R.~G., et al.\ 2012, \aap, 548, A119. doi:10.1051/0004-6361/201219505

\bibitem[Cutri et al.(2013)]{Cutri et al.(2013)} Cutri, R.~M., Wright, E.~L., Conrow, T., et al.\ 2013, Explanatory Supplement to the AllWISE Data Release Products, by R. M. Cutri et al.

\bibitem[Eisenhardt(2019)]{Eisenhardt(2019)} Eisenhardt, P.\ 2019, American Astronomical Society Meeting Abstracts \#233

\bibitem[Hambly et al.(2008)]{Hambly et al. (2008)} Hambly, N.~C., Collins, R.~S., Cross, N.~J.~G., et al.\ 2008, \mnras, 384, 637. doi:10.1111/j.1365-2966.2007.12700.x

\bibitem[Kirkpatrick et al.(2011)]{Kirkpatrick et al. 2011} Kirkpatrick, J.~D., Cushing, M.~C., Gelino, C.~R., et al.\ 2011, \apjs, 197, 19. doi:10.1088/0067-0049/197/2/19

\bibitem[Kirkpatrick et al.(2016)]{Kirkpatrick et al. 2016} Kirkpatrick, J.~D., Kellogg, K., Schneider, A.~C., et al.\ 2016, \apjs, 224, 36. doi:10.3847/0067-0049/224/2/36

\bibitem[Kirkpatrick et al.(2021a)]{Kirkpatrick et al. 2021a} Kirkpatrick, J.~D., Gelino, C.~R., Faherty, J.~K., et al.\ 2021, \apjs, 253, 7. doi:10.3847/1538-4365/abd107

\bibitem[Kirkpatrick et al.(2021b)]{Kirkpatrick et al. 2021b} Kirkpatrick, J.~D., Marocco, F., Caselden, D., et al.\ 2021, \apjl, 915, L6. doi:10.3847/2041-8213/ac0437

\bibitem[Kuchner et al.(2017)]{Kuchner et al. 2017} Kuchner, M.~J., Faherty, J.~K., Schneider, A.~C., et al.\ 2017, \apjl, 841, L19. doi:10.3847/2041-8213/aa7200

\bibitem[Lang(2014)]{Lang(2014)} Lang, D.\ 2014, \aj, 147, 108. doi:10.1088/0004-6256/147/5/108

\bibitem[Lawrence et al.(2007)]{Lawrence et al.(2007)} Lawrence, A., Warren, S.~J., Almaini, O., et al.\ 2007, \mnras, 379, 1599. doi:10.1111/j.1365-2966.2007.12040.x

\bibitem[Lucas et al.(2008)]{Lucas et al.(2008)} Lucas, P.~W., Hoare, M.~G., Longmore, A., et al.\ 2008, \mnras, 391, 136. doi:10.1111/j.1365-2966.2008.13924.x

\bibitem[Mace et al.(2013)]{Mace et al. 2013} Mace, G.~N., Kirkpatrick, J.~D., Cushing, M.~C., et al.\ 2013, \apj, 777, 36. doi:10.1088/0004-637X/777/1/36

\bibitem[Mainzer et al.(2011)]{Mainzer et al. 2011} Mainzer, A., Bauer, J., Grav, T., et al.\ 2011, \apj, 731, 53. doi:10.1088/0004-637X/731/1/53

\bibitem[Marocco et al.(2021)]{Marocco et al. 2021} Marocco, F., Eisenhardt, P.~R.~M., Fowler, J.~W., et al.\ 2021, \apjs, 253, 8. doi:10.3847/1538-4365/abd805

\bibitem[Meisner et al.(2018)]{Meisner et al. (2018)} Meisner, A.~M., Lang, D., \& Schlegel, D.~J.\ 2018, \aj, 156, 69. doi:10.3847/1538-3881/aacbcd

\bibitem[Meisner et al.(2020)]{Meisner et al. 2020} Meisner, A.~M., Caselden, D., Kirkpatrick, J.~D., et al.\ 2020, \apj, 889, 74. doi:10.3847/1538-4357/ab6215

\bibitem[Meisner et al.(2021)]{Meisner et al. 2021} Meisner, A.~M., et al.\ 2021, \apj, in press.

\bibitem[Pinfield et al.(2014)]{Pinfield et al. 2014} Pinfield, D.~J., Gomes, J., Day-Jones, A.~C., et al.\ 2014, \mnras, 437, 1009. doi:10.1093/mnras/stt1437

\bibitem[Schlafly et al.(2019)]{Schlafly et al.(2019)} Schlafly, E.~F., Meisner, A.~M., \& Green, G.~M.\ 2019, \apjs, 240, 30. doi:10.3847/1538-4365/aafbea

\bibitem[Schneider et al.(2020)]{Schneider et al. 2020} Schneider, A.~C., Burgasser, A.~J., Gerasimov, R., et al.\ 2020, \apj, 898, 77. doi:10.3847/1538-4357/ab9a40

\bibitem[Skrutskie et al.(2006)]{Skrutskie et al. 2006} Skrutskie, M.~F., Cutri, R.~M., Stiening, R., et al.\ 2006, \aj, 131, 1163. doi:10.1086/498708

\bibitem[Wright et al.(2010)]{Wright et al. 2010} Wright, E.~L., Eisenhardt, P.~R.~M., Mainzer, A.~K., et al.\ 2010, \aj, 140, 1868. doi:10.1088/0004-6256/140/6/1868

\bibitem[Zhang et al.(2017)]{Zhang et al. 2017} Zhang, Z.~H., Pinfield, D.~J., G{\'a}lvez-Ortiz, M.~C., et al.\ 2017, \mnras, 464, 3040. doi:10.1093/mnras/stw2438

\bibitem[Zhang et al.(2019)]{Zhang et al. 2019} Zhang, Z.~H., Burgasser, A.~J., G{\'a}lvez-Ortiz, M.~C., et al.\ 2019, \mnras, 486, 1260. doi:10.1093/mnras/stz777

\end{thebibliography}
\end{document}